\documentclass[12pt]{iopart}
\usepackage{graphicx}
\usepackage{graphics}
\usepackage{dcolumn}
\usepackage{bm}
\usepackage{pstricks}
\usepackage{amssymb}
\usepackage{hyperref}
\usepackage{subeqn}

\definecolor{lineadecolor}{rgb}{0.35,0.5,0.6}
\definecolor{ddcol}{rgb}{0.8,0.1,0.1}
\definecolor{subsectioncolor}{rgb}{0.1,0.01,0.5}
\definecolor{celeste}{rgb}{0.8,0.87,0.99}

\def\nn{\nonumber}
\newcommand{\ba}{\begin{eqnarray}}
\newcommand{\ea}{\end{eqnarray}}
\def\be{\begin{equation}}
\def\ee{\end{equation}}

\def\spin{\vec{\mathbf{S}}}

\begin{document}

\title{Phase diagram study of a dimerized spin-S zig-zag ladder}

\author{ J.M.\ Matera}
 \address{IFLP - CONICET. Departamento de F\'isica, Facultad de Ciencias Exactas. Universidad Nacional de La Plata,
C.C.\ 67, 1900 La Plata, Argentina.}

\author{ C.A.\ Lamas}
\address{IFLP - CONICET. Departamento de F\'isica, Facultad de Ciencias Exactas. Universidad Nacional de La Plata,
C.C.\ 67, 1900 La Plata, Argentina.}

\begin{abstract}

The phase diagram of a frustrated spin-$S$ zig-zag ladder is studied through different  numerical and analytical methods. 
We show that for arbitrary $S$, there is a family of Hamiltonians for which a fully-dimerized state is an  exact ground state,
being the Majumdar-Ghosh point a particular member of the family.
We show that the system presents a transition between a dimerized phase to a N\'eel-like phase for $S=1/2$, and
spiral phases can appear for large $S$.
The phase diagram is characterized by means of a generalization of the usual Mean Field Approximation (MFA).
The novelty in the present implementation is to consider the strongest coupled sites as the unit cell. The gap and the excitation spectrum is analyzed through the Random Phase Approximation (RPA).  Also, a perturbative treatment to obtain the critical points is discussed.  Comparisons of the
 results with numerical methods like DMRG are also presented.
\end{abstract}
\maketitle


\section{Introduction}

In physics, exact results have been proved to be extremely useful both from a conceptual point of view as well as for 
practical reasons, as  references to check approximated methods that we use
to solve more realistic models, for which the exact solution is not available.
 For example, for $S=1$, the exact ground state (GS) of the AKLT model \cite{AKLT} has been very important in the confirmation 
of Haldane's prediction \cite{Haldane}.
In this context, quantum spin ladders (QSL) represent a special scenario to obtain analytical results and have played an important role in
the past few years\cite{Fisher,ladders_L,zigzag_C1,Lamas-Ralko_2013,Hida-Affleck_2005,OYA,WFNK.13,MV.12,MMM.13}.
These systems are interesting from the theoretical point of view as examples of one dimensional correlated quantum systems,
that can be used to study quantum phase transitions related to the existence of a spin gap.
 The case of the two-leg zig-zag ladder is one of the most famous examples of  frustrated spin models and highlights the role played by
 frustration, describing a number of quasi-one  dimensional compounds like Cs$_2$CuCl$_4$\cite{cscucl}, KCuCl$_3$\cite{TlCuCl}, TlCuCl$_3$\cite{TlCuCl}
 and NH$_4$CuCl$_3$\cite{NhCuCl}. For this reason, the ladders with zig-zag couplings between quantum spin chains has received much attention
\cite{Chitra,Allen,White,Affleck1,Nersesyan}.

In the zig-zag ladder, frustration reduces anti-ferromagnetic correlations and the tendency towards N\'eel order, which would lead to 
a dimerized phase.
 A particularly useful example of an exact result in QSL corresponds to the exact GS of the  Majumdar-Ghosh Model
 (MGM)\cite{MG}, consisting on a  translational, $SU(2)$ invariant spin $S=1/2$ chain with a particular relation between
 the  first and second neighbor couplings. This model presents a degenerated ground state, spanned by two  non-orthogonal fully-dimerized states  (i.e. a state that can be written as a product of singlet
 states between contiguous sites).  

The aim of this work is to generalize the result for the MGM in two ways. On the one hand, we show 
that there is a larger family on the two-legs zig-zag frustrated Quantum Spin Ladder (QSL) which also presents as its exact GS a fully-dimerized state. 
On the other hand, we show that this region exists also for systems with larger local spin $S$.
Then, we discuss how the large $S$ limit arise on this model, washing up the dimerized phase and
recovering the N\'eel order, predicted by a semi-classical expansion. \cite{MV.12,WFNK.13,MMM.13}

The paper is organized as follows. In the next section, we discuss the details of the model and  we show,
for every  value of the local spin $S$, the existence of a fully-dimerized ground state along a continuous 
line in the parameter space.
In section \ref{sec:mft}  we show, by means the usual variational MFA based on individual spins, how to obtain an upper bound to the region where 
the fully-dimerized state is the true ground state of the system. Also we present a generalization to MFA which  captures the main features
 of the dimerized phase. 
 In section \ref{sec:gap} analytical expressions for the gap, the elementary excitations of the model as well as
 the localization  of the boundaries of the dimerized region are obtained through the Random Phase Approximation (RPA) formalism developed on previous works \cite{MRC.08,MRC.10}.
 Then, a discussion of the limitations of the technique and how they can be overcome through perturbative corrections is presented.
 Section \ref{sec:conclusions} is devoted to conclusions and perspectives.
\section{Special features of the model}
\label{sec:model}
\begin{figure}
\begin{centering}
\includegraphics[width=0.7\columnwidth]{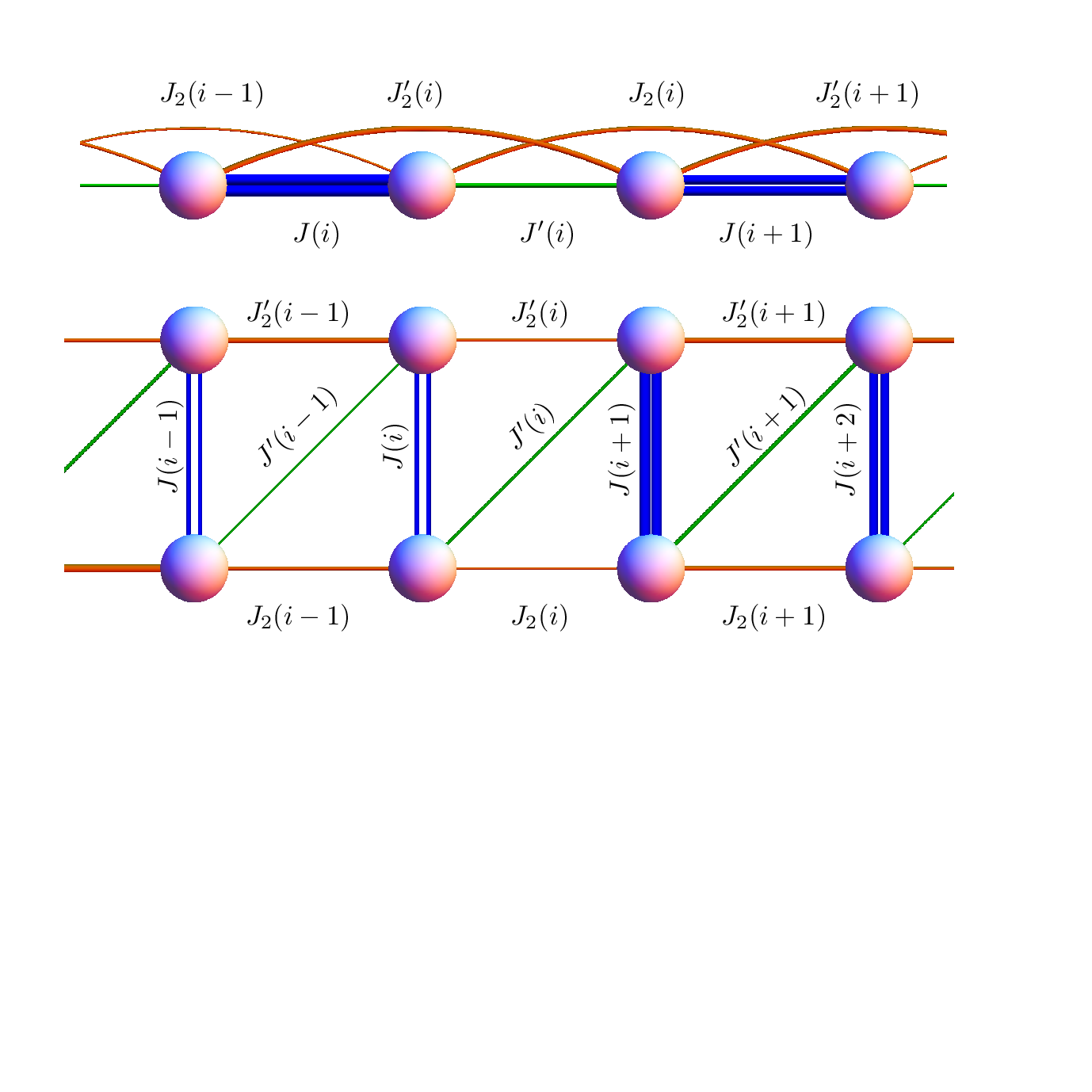}
\vspace*{-.3cm}
\caption{(Color Online) Two different cartoons for the zig-zag ladder. Top: 1D chain with second neighbors couplings. Bottom: ladder with zig-zag couplings.}
\label{fig:laddera}
\end{centering}
\end{figure}

We consider the following Heisenberg model on a two legs spin-$S$ zig-zag ladder
\ba
\label{eq:general_Hamiltonian}
\nn
H&=&\sum_{i=1}^{N}J(i) \spin_{2i-1}\cdot\spin_{2i}+J'(i) \spin_{2i}\cdot\spin_{2 i+1}\\
&+& J_{2}(i) \spin_{2i}\cdot\spin_{2(i+1)}+ J^{'}_{2}(i)\spin_{2i-1}\cdot\spin_{2(i+1)-1}
\ea
which also can be seen as a spin chain  with next-nearest neighbors (NNN) interactions (See Fig \ref{fig:laddera}).
In Eq. (\ref{eq:general_Hamiltonian}), $\spin_{i}$ is a vectorial spin-S operator  on site $i$ with components
${\bf S}^x_{i}$, ${\bf S}^y_{i}$ and ${\bf S}^z_{i}$;
$N$ represents the number of ``rungs'' on the ladder and $J(i)$, $J'(i)$, $J_{2}(i)$ and $J'_{2}(i)$ are bond dependent coupling constants as they are depicted in
Fig. \ref{fig:laddera} . As we are interested on the frustrated regime, we will consider only the case where all the couplings are
positive (all the interactions are anti-ferromagnetic). Note that this model includes the case of systems which breaks the translational symmetry, including the case with ``disordered couplings''

This family of Hamiltonians interpolates among several well known systems. If $J'(i)$ or $J_{2}(i)$ and $J_{2}'(i)$ vanish, the lattice is non frustrated and it can be analyzed by different means. 
In particular, for $J'(i)=J(i)=J$ ($J$ being a site independent value) and $\forall _i$ $J_2(i)\equiv 0$, the system is reduced to an isotropic Heisenberg chain
which can be solved exactly for spin $1/2$ using the Bethe ansatz\cite{Bethe}, and which has been extensively studied \cite{OYA,spin_chain_1}. The GS of this systems
 presents quasi long range order (QLRO) correlations, related to the logarithmic
 violation of area laws on its entanglement entropy\cite{ECP.10,DET.11}.

On the other hand, if $\forall _{i}$ $J'(i)=0$ and  $J_2(i)=J(i)=J$, the system is reduced to a Heisenberg two-leg ladder.
Taking $J/J_{2} \rightarrow 0 $ we have two decoupled  Heisenberg chains,
for which at $S=1/2$ are gapless. It was predicted \cite{Dagotto_1993}  that for $S$
half-integer and $J'=0$ the spin-gap would vanish only at $J/J_{2} = 0$.
Then, for half-integer local spin, the ladder would always be in gaped regime in contrast to
the more familiar case of the Nearest Neighbor Heisenberg model (corresponding to $J'=J$ and $J_2=0$) which is gapless.

Interestingly, in the special case where $J_{2}(i)+ J^{'}_{2}(i)=J'(i)$, we can determine the GS of the system 
for a given range of $0\leq J_{2}(i),J_{2}^{'}(i) <J'(i) < J_{c}^{'}$, where the upper bound $J_{c}^{'}$ depends of the spin strength. 
For $S=\frac12$ and $2 J_{2}^{'}(i)=2J_2(i)=J'(i)=J(i)=J$ this GS becomes degenerate and corresponds to the GS
of the MGM\cite{MG}.
\subsection{Dimerized ground state}
If $\forall _i$ $0<J(i),J'(i),J_2(i),J^{'}_2(i)$, the system becomes frustrated, which in general implies a complex structure of
the GS and its excitation. In this case some of the usual techniques used 
to deal with this kind of problems may fail at zero temperature.
However, for  $J'(i)=J_2(i)+J_2^{'}(i)<J(i)$, the Hamiltonian (\ref{eq:general_Hamiltonian}) presents a factorized eigenstate which 
corresponds to the GS for $J'S\ll \min_i J(i)$. To see it, we can rewrite the Hamiltonian in terms of
local operators on each rung:
\begin{eqnarray}
\nonumber
{\bf H}&=& - S(S+1)\,\sum_iJ_i+ \\ 
\nonumber && + \sum_{i} J_i\frac{{\bf J}^2_{i}}{2}  
+ \sum_i \frac{J_2(i)+J^{'}_2(i)+J'(i)}{4} \vec{\bf J}_i\cdot \vec{\bf J}_{i+1}+\\
&& \nonumber +\sum_i \frac{J'(i)+J_2(i)-J_2'(i)}{4} \vec{\bf K}_i\cdot \vec{\bf J}_{i+1}- \\
&&\nonumber  -\sum_i  \frac{J'(i)+J'_2(i)-J_2(i)}{4} \vec{\bf J}_i\cdot \vec{\bf K}_{i+1}+\\
&& 
+ \sum_i  \frac{J_2(i)+J_2^{'}(i)-J'(i)}{4}\vec{\bf K}_i\cdot \vec{\bf K}_{i+1}     \label{hjk}
\end{eqnarray}
where
\begin{eqnarray}
\vec{\bf J}_{i}&=&\vec{\bf S}_{2i}+\vec{\bf S}_{ 2i-1}\label{eq:JopDef} \\
\vec{\bf K}_{i}&=&\vec{\bf S}_{2i}-\vec{\bf S}_{2i-1} \label{eq:KopDef}\,.
\end{eqnarray}
$\vec{\bf J}_i$ is the total angular momentum of the rung $i$,
and $\vec{\bf K}_i$ is another set of local vectorial observable which completes the full local 
Lie algebra relevant to the problem. Its components close the following Lie algebra:
\begin{subequations}
\label{eq:localliealgebra}
\begin{eqnarray}
   \left[{\bf J}_{\mu},{\bf J}_{\nu}\right]&=& {\bf i} \epsilon_{\mu\nu\eta} {\bf J}_{\eta}\\
   \left[{\bf J}_{\mu},{\bf K}_{\nu}\right]&=& {\bf i} \epsilon_{\mu\nu\eta} {\bf K}_{\eta}\\
   \left[{\bf K}_{\mu},{\bf K}_{\nu}\right]&=& {\bf i} \epsilon_{\mu\nu\eta} {\bf J}_{\eta}
\end{eqnarray}
\end{subequations}
where $\epsilon_{\mu\nu\eta}$ is the fully antisymmetric Levi-Civita symbol 
and ${\bf i}=\sqrt{-1}$ is the imaginary unit.
If we set the constraints
\begin{equation}
  \label{eq:constrMG}
  J'(i)=J_2(i)+J'_2(i)\,,
\end{equation}
the last term in (\ref{hjk}) vanishes and the state
\begin{eqnarray}
\label{eq:dimerstate}
  |{\rm dimer}\rangle&=&\bigotimes_{i=1}^{N}|{\rm singlet}\rangle \hspace{1cm}\mbox{with}\\
  \label{eq:sigletS}
 |{\rm singlet}\rangle&=& \frac{1}{\sqrt{2 S+1}}\sum_{m=-S}^{S}(-1)^{m+S}|-m\rangle|m\rangle
\end{eqnarray}
results an eigenstate of ${\bf H}$ with energy $E_{\rm dim}=-J N S(S+1)$, due to  $\vec{\bf J}_i|{\rm singlet}\rangle_i=0$. Noteworthy, this result is valid for any value of the local spin magnitude $S$.  

The state (\ref{eq:dimerstate}) has not a classical analog in terms of individual classical magnetic moments, corresponding to a phase with a characteristic quantum behavior: a dimerized phase.
In this phase, despite the spin-spin correlations vanish between non continuous sites, there could exist correlations between pairs of its elementary excitations.
Another characteristic of this order is that there is a spin gap which prevents the occurrence of a spontaneous breaking in the global $SU(2)$ symmetry.
However, the determination of the gap can be a non-simple task. Although the dimerized GS may remains stable when we change the couplings maintaining the constraint (\ref{eq:constrMG}), the excited states strongly depend on the values of the local couplings.
\begin{figure}
\begin{center}
\includegraphics[width=0.8\columnwidth]{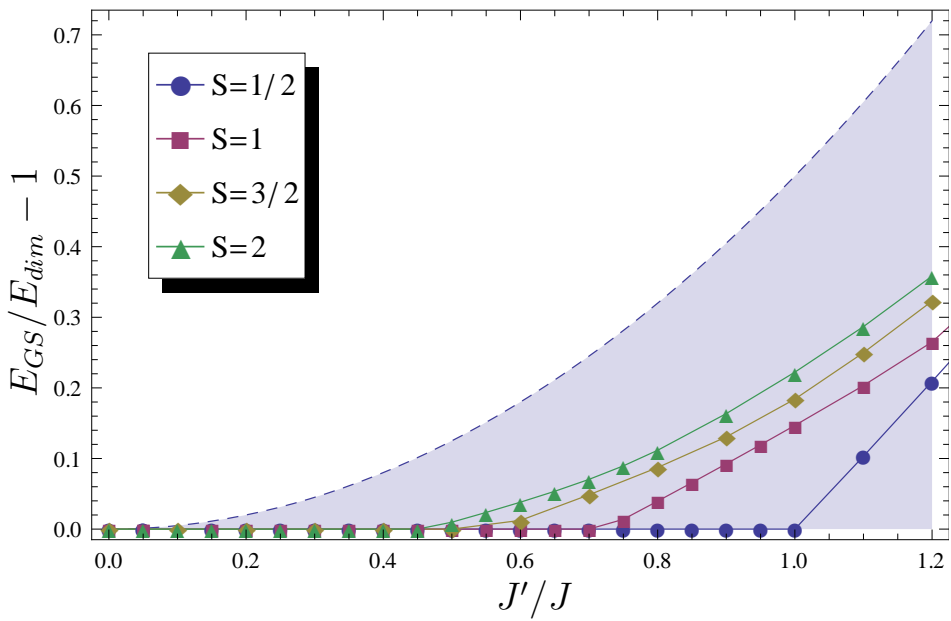}\\
\caption{(Color Online) GS energy relative to the energy of the dimerized state $E_{\rm GS}/E_{\rm dim}-1$
as a function of $J'/J$ over the line $J_2=J'/2$ for different spin values in a $2\,N=40$ sites system. The results was calculated through the 
DMRG method (see  \ref{sec:nummeth}). The border of the shaded region (dashed line) corresponds to the mean field result  (\ref{eq:mfsepfactline}). }
\label{sasa}
\end{center}
\end{figure}

The few systems where an exact GS can be obtained have played a very important role in quantum magnetism. For example, 
we can cite the exact GS of the AKLT point \cite{AKLT} which has been very important in the confirmation 
of Haldane's prediction \cite{Haldane} that the spectrum of an integer-S spin chains is gaped.
In the present model, the dimerized eigenstate is surprisingly robust. 
For example we can change the couplings $J(i)$, $J'(i)$ and $J_2(i)$ in Hamiltonian (\ref{eq:general_Hamiltonian}) 
and, preserving the constraints (\ref{eq:constrMG}), the dimerized state is still an eigenstate;
this is a very important property inside the regions where this eigenstate is the GS of the system:
since excited states are strongly dependent of the distribution of couplings, we can change 
the distribution of couplings in order to reduce the gap.
A similar property of the AKLT model can be exploited for spin 1. In fact a 
very interesting question is if it there exists some coupling distribution that makes the system gapless.
 The advantage of this system is that its robust GS is present over a extended region of the very
 large parameter space for any value of the local spin.
Since the excited states in the general case can be very complicated, a such study of the many possible excited states deserves a separate publication. In order to obtain a qualitative picture,
 in the present manuscript we focus mainly on the  restricted case of the uniform ladder $J(i)=J$, $J'(i)=J'$ and $J_2(i)=J'_2(i)=J_2$, where $J$, $J'$ and $J_{2}$ are positive and site-independent numbers. Besides, to avoid boundary effects,  we will restrict to the case of periodic boundary conditions and $N$ even.
 Under these assumptions, we can easily show that the $|{\rm dimer}\rangle$ state in Equation (\ref{eq:sigletS}) is a true GS of the system if the condition
\begin{equation}
  \label{eq:exdimcond}
J'=2J_2<\left\{
  \begin{array}{c}
J\hspace{.5cm} S=\frac{1}{2}\\
\frac{J}{S+1}\hspace{.5cm} S\geq 1
 \end{array}\right.
\end{equation}
is provided. To see this, we can write the Hamiltonian (\ref{eq:general_Hamiltonian}) in a more convenient way:
\ba
\nn
{\bf H}=\frac{1}{2}\sum_{i=1}^{2N}{\bf h}_i
\ea
with
\ba
\nn
{\bf h}_i&=& \frac{J'}{2} \left[  (\spin_{i-1}+\spin_{i}+\spin_{i+1})^{2}
-\spin_{i}^{2}-\spin_{i-1}^{2}-\spin_{i+1}^{2}\right] + \\
&+&\frac{\delta J}{2}\left[  (\spin_{i}+\spin_{i-(-1)^{i}})^{2}
-\spin_{i}^{2}-\spin_{i-(-1)^{i}}^{2}\right]
\ea
where $\delta J=J-J'$. From this decomposition we can see that the GS energy per rung is 
{\small
$$\frac{E_{\rm GS}}{N}\geq \displaystyle{\min_{l\leq 2S,\; l\in \mathbb{Z}}} \frac{J'}{2} (|S-l|)(|S-l|+1)+\frac{\delta J}{2} l(l+1)- (J +\frac{J'}{2})  S(S+1),$$
}
where the right hand side corresponds to the minimum eigenvalue of all the ${\bf h}_i$ terms.
When the condition (\ref{eq:exdimcond}) is satisfied, this lower bound coincides with the energy
associated to the state $|{\rm dimer}\rangle$ and hence, it is a GS.
Moreover, due to for $J>J'$ there is not any other state which minimizes all the ${\bf h}_i$ at a time and
hence, this GS results non-degenerated.
Equation (\ref{eq:exdimcond}) represents a sufficient condition for the existence of the dimerized
GS, but the true range of couplings where this state is a real GS can be larger.
In Figure \ref{sasa} the relative energy $E_{\rm GS}/E_{\rm dim}-1$ as a function of $J'/J$  
over the line $J_{2}=\frac{J'}{2}$ for a $2\,N=40$ sites system with different local spin values is depicted.
This quantity vanishes when the energy of the system is equal to the energy of the dimerized eigenstate.
We can see that for a given spin  $S$, the values of $J'$ where $E_{\rm GS}/E_{\rm dim}-1$ becomes different from zero are larger than the
values predicted in Eq. (\ref{eq:exdimcond}).
Dashed line corresponds to the single site mean field result discussed in the next Section, corresponding
to the large $S$ limit. In Section \ref{sec:gap} we show that for large $S$ the next to leading order 
is proportional to $S^{-1}$. 
In the next section we will show also that if $J'/J>\sqrt{2/S}$ there is a fully factorized state
(the mean field state) which has energy lower than the dimerized eigenstate. This allow us to 
obtain an upper bound value for the transition coupling $J_{c}(S)$, whereas Eq. (\ref{eq:exdimcond})
represents a lower bound. 

\begin{figure}

\begin{center}
\includegraphics[width=0.65\columnwidth]{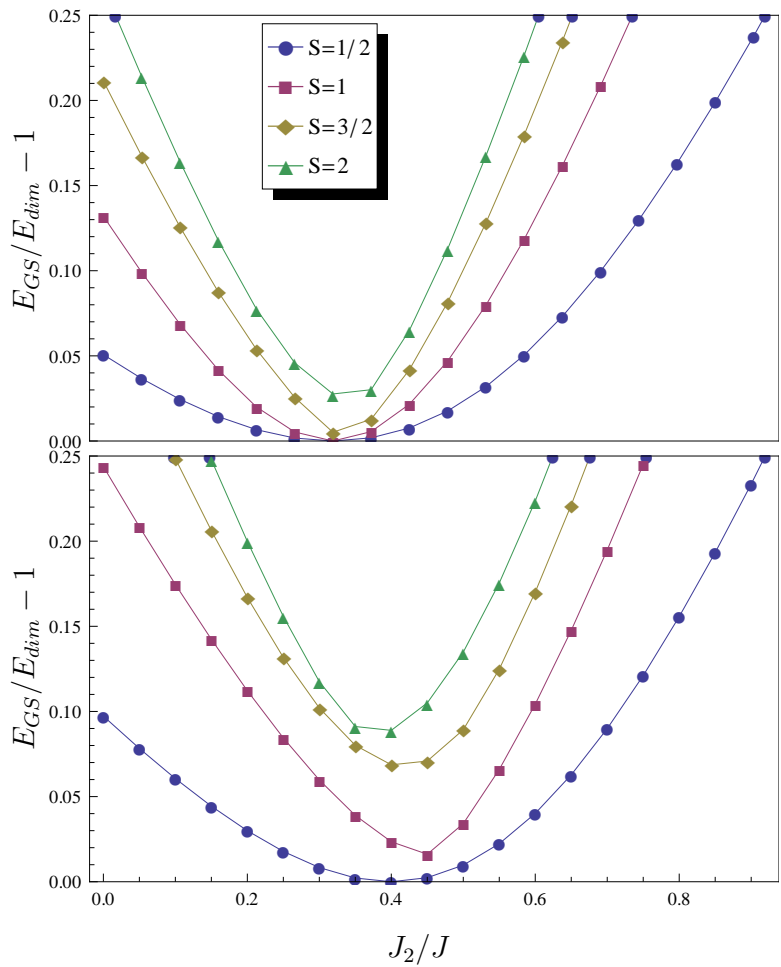}\\
\caption{(Color Online) GS energy relative to the dimerized energy  $E_{\rm dim}$ as a function of $J_2$ for different 
values spin values for $J'=0.6 J$ (Top panel) and $J'=0.8J$ (Bottom panel), calculated through DMRG in a $2\,N=40$
sites system. On the top panel we can appreciate how at the line $J'=2 J_2$
the energy corresponds to the dimerized phase for $S=1/2$ and $S=1$, but results lower (higher values of the relative energy) for $S=3/2$ and $S=2$.
  On the Bottom panel, the results for $J'=0.8 J$ are depicted. For this value, the dimerization is only reached
  for $S=1/2$. Besides, we can observe that the deviation from the energy of the dimerized phase grows up with the
  local spin magnitude and with the value of $J'$.  }
\label{fig:evsJ2}
\end{center}
\end{figure}

 In Figure \ref{fig:evsJ2} we show the relative energy calculated by
DMRG as a function of $J_{2}$ for different values of $S$. Upper panel corresponds to $J'=0.6$ where the systems with 
$S=1/2$ and $S=1$ reach the fully-dimerized GS at the $J_2=J'/2=0.3 J$ point. We can observe in the figure 
that, for this two cases, the 
curves come together at that point,  whereas the curves corresponding to $S=3/2$ and $S=2$ have a lower energy (which corresponds to a larger relative energy). 
Lower panel in Fig. \ref{fig:evsJ2} corresponds to $J'=0.8$ where just the $S=1/2$ system reaches the dimerized 
GS.

 \section{Mean Field Treatment}
\label{sec:mft}
One of the most elementary but versatile approach to approximate the GS of interacting systems is
through the well known mean field approximation.
This approximation can be expressed in terms of a variational approach over the family of product states
of each individual component of the system. Essentially, it consists on neglecting correlations between 
fluctuations associated to the  different components. The resulting state can be considered classical in several ways.
To begin with, as it does not present entanglement between components, it does not support quantum correlations between components.
Another reason to consider it as a ``classical limit'' is related to the particular form of the interacting terms in typical Hamiltonians, which consist of
bilinear forms on the  generators of the local algebras.
In particular, for spin systems, it leads to solutions of the mean field problem in the form of
a product of spin coherent states. 
These can be represented, in the same way that  classical magnetic moments, as a set of arrows of fixed length,
each one on a different site. Besides, the optimum state coincides with the optimum self-consistent configuration of the classical model.
erv

Now, suppose that we try to approximate the GS over the fully-dimerized region with such product state. 
There, the  exact GS is a product of consecutive  singlet states. For the spin $S$ ladder, the maximum overlap of this
state with a product state is $1/(\sqrt{2 S+1})^N$, attained for a N\'eel-type state 
$|{\rm Neel}\rangle=\bigotimes_{i=1}^N(|\uparrow\rangle|\downarrow\rangle)$.
In this way, when the system is in a true dimerized phase, there is no classical state which gives a faithful
representation of the dimerized state: moreover, the overlap with the exact state vanishes exponentially with
the size of the system.

This is not surprising, since it is known that mean field approaches are not extremely reliable in dealing with 
quasi-one-dimensional systems. On the other hand, if we observe that for a certain region the energy associated to 
the MF state is lower than the fully-dimerized state, we will have shown that on that region the fully-dimerized state is 
not the true ground state. This fact allow us in the next paragraphs to establish an upper bound to the interval of 
values of the coupling constants where the fully-dimerized state is the true GS of the system. Hence, as we will see 
later, the classical limit comes in an unexpected way, through a level crossover which eliminates the dimerized phase 
for large enough $J'$ or $S$.
In the remainder of this section we  discuss how to arrive to analytical expressions of the ``classical limit'' solution, and its particularities.
Besides, we propose a way to generalize the mean field solution in a suitable way to deal with dimerized phases which becomes \emph{exact} over the fully-dimerized line.
Then, at the end of this section, we come back to the question of the classical limit and how the classical solution appears.
The issue of the stability of both kind of solutions are discussed in the next section.  
\subsection{``Classical'' ($S\rightarrow \infty$) phase diagram}
\begin{figure}
\begin{centering}
\includegraphics[width=0.5\columnwidth]{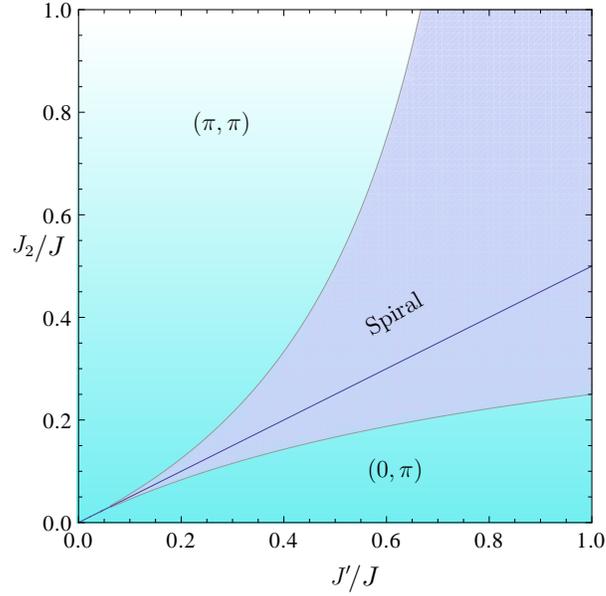}\\
\vspace*{1cm}\hspace*{.5cm}
\includegraphics[width=0.5\columnwidth]{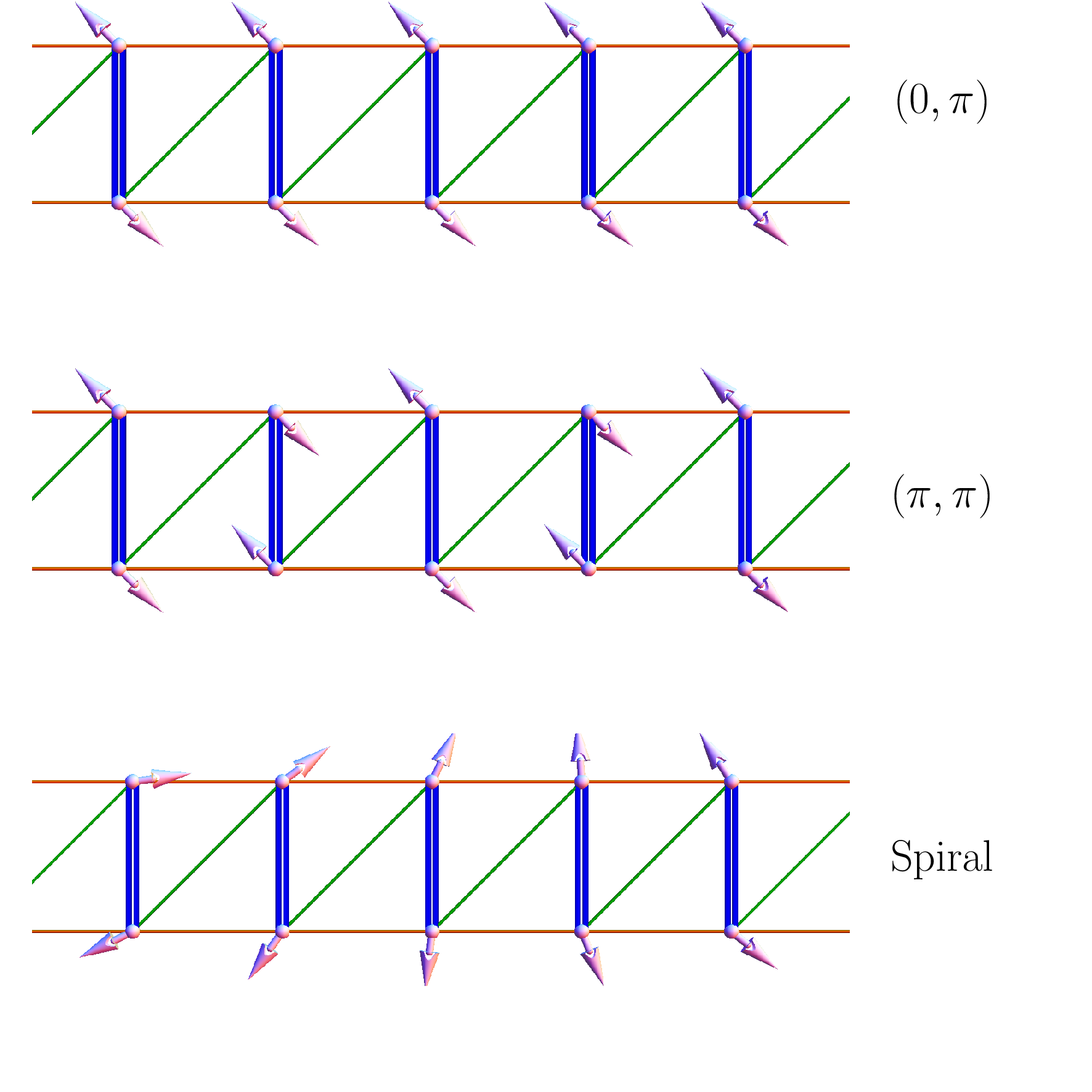}
\vspace*{-1.cm}
\caption{(Color Online) Top: Phase diagram corresponding to the classical limit. 
The gray region corresponds to the spiral phase, 
the regions $(\pi,\pi)$ and $(0,\pi)$ correspond to colinear phases.
Bottom: Cartoon of the classical configurations corresponding to the regions in the phase diagram of the bottom panel.
}
\label{fig:classicalphase_diag}
\end{centering}
\end{figure}

The GS of the ``classical'' model can be known by studying a unit cell, as composed by two magnetic moments with fixed magnitude. This approach is equivalent to look for an variational approximation over the family of product states of fully polarized spin states
\begin{equation}
  \left|\rm MF\right\rangle_{\rm classical}= \bigotimes_{i=1}^{N} |\vec{s}_{2\,i-1}\rangle|\vec{s}_{2\,i}\rangle \label{mfss1}\,.
\end{equation}
Besides, because the Hamiltonian that we are considering is \emph{linear} on each local spin projection operators,
this approach is completely equivalent to a full mean field treatment. In this way, the problem is reduced to find the angles determining the directions of all local spins, which can be found 
 by numerical methods, for instance, in a self-consistent way. To allow us to go forward in an analytical description, we can reduce the problem by looking for
 solutions over a subfamily with certain symmetries. 
 A first reasonable assumption is that the states that we are looking for are symmetrical under a reflection with
 respect to the plane $xz$, which is a symmetry of the Hamiltonian. 
 For this reason, we can expect that there were a solution with  all the states $|\vec{s}_{i}\rangle$ polarized in that
 plane\footnote{For finite systems, a numerical analysis shows that in general, the energy can be lowered if we allow the magnetic
 moments to move sightly away from the plane.}.
A second assumption is that $\forall_i$ the angle between $\vec{s}_k$ and $\vec{s}_{k+2}$ is constant, due to the global translational symmetry.
The resulting family corresponds to the planar spiral states

\begin{eqnarray}
  \label{eq:plspiralstates}
  |\theta\phi\rangle_{spiral}&=&\bigotimes_{j=1}^{2N}|\varphi_j\rangle \,,
\hspace{.4cm}  \varphi_{j+1}=\theta\,j -(-1)^j\frac{\phi}{2} \hspace{.5cm}
\end{eqnarray}
being $|\varphi \rangle=\exp(-{\bf i} \varphi {\bf S}^{y})|\rightarrow\rangle$ the local coherent state polarized along the direction $\vec{s}=(\cos(\varphi),0,\sin(\varphi))$. The energy $E_{\rm sep}[\theta,\phi]=\langle \theta \phi|{\bf H}|\theta\phi\rangle$ is hence given by
\ba
\frac{E_{\rm sep}(\theta,\phi)}{NS^2}\!=\!J\cos\phi+J' \cos(\theta\!-\!\phi)+2 J_{2}(\cos\theta)\,. \label{mffenergy}
\ea
Looking for the extremes of this function we obtain the following energies
\begin{subequations}
\label{eq:mfsepenergies}
\ba
\frac{E_{\rm sep}(0,\pi)}{NS^2}&=&2J_{2}-J-J'\\
\frac{E_{\rm sep}(\pi,0)}{NS^2}&=&J-J'-2J_{2}\\
\frac{E_{\rm sep}(\pi,\pi)}{NS^2}&=&J'-J-2J_{2}\\
\frac{E_{\rm sep}(\tilde{\theta},\tilde{\phi})}{NS^2}&=&J\cos\tilde{\phi}+J' \cos(\tilde{\theta}-\tilde{\phi})+ 2 J_{2} \cos\tilde{\theta}
\ea
\end{subequations}
where
\begin{subequations}
\label{eq:mfsepenangl}
\ba
\cos{\tilde{\theta}}&=&\frac{J J'}{8 {J_2}^2}-\frac{J}{2 J'}-\frac{J'}{2 J}\\
\cos{\tilde{\phi}}&=&\frac{J_{2} J'}{J^2}-\frac{J_{2}}{J'}-\frac{J'}{4 J_{2}}.
\ea
\end{subequations}
Phases corresponding to energies $E_{\rm sep}(\pi,\pi)$ and $E_{\rm sep}(0,\pi)$ are labeled as $(\pi,\pi)$
and $(0,\pi)$ respectively, and corresponds to two different N\'eel orders.
The phase with energy $E_{\rm sep}(\tilde{\theta},\tilde{\phi})$ is a spiral phase (see Fig. \ref{fig:classicalphase_diag} ).
 
If we consider now the case of $J'=2J_2$, $\cos(\tilde{\theta})\rightarrow -\frac{J'}{2 J}$,  $\tilde{\phi}\rightarrow 2 (\pi-\tilde{\theta})$  and the energy
 results
\begin{equation}
  \label{eq:mfsepfactline}
E_{\rm sep}\rightarrow -J N S^2 \left(1+ \frac{1}{2}\left(\frac{J'}{J}\right)^2\right)\,.  
\end{equation}

We can make two observations about it. On the one hand, we can see that for $(J'/J)^2 >\frac{2}{S}$ the energy of this mean field  is lower than the
associated to the dimerized state, which suggests to identify the ``classical limit'' with the ``large $S$ limit''. On the other hand, for small spin, we can observe
 that this state is a very rough approximation to the true GS in several ways: First of all, for $S\leq 1$,
$\frac{E_{\rm sep}-E_{\rm GS}}{E_{\rm GS}}\approx 1$, in a way that does not allow us to consider the difference as an small correction over the mean field result.
Other symptom of the in-accurateness of the approximation is related to the lack of similarity between the true GS and the optimum spiral state. In 
order to quantify this difference, we can compute the \emph{Fidelity} \cite{NC.00} between the density matrix of any pair of spins. The fidelity 
is defined as
\begin{equation}
  {\cal F}[\rho_1,\rho_2]={\rm Tr}\sqrt{\sqrt{\rho_2} \rho_1 \sqrt{\rho_2} }
\end{equation}
(see \ref{sec:propfidelity}) where $\rho_{1,2}$ are two different density matrices. For pure states ($\rho_i=|\psi_i \rangle\langle \psi_i|$), this quantity is 
reduced to the absolute value of the overlap between both states
$ {\cal F}[|\psi_1\rangle|\psi_2\rangle]=|\langle \psi_1|\psi_2\rangle|$.
In the dimer state, the density matrix of any pair of spins is given by
$$
\rho^{\rm(dimer)}_{ij}= \left \{ 
  \begin{array}{l  r}
    |{\rm singlet}\rangle\langle {\rm singlet} |   & \mbox{in the same rung}    \\
    \frac{{\bf 1}}{(2s+1)^2} & \mbox{otherwise}. 
  \end{array}
\right.
$$
while for the spiral state
$$
\rho^{\rm(spiral)}_{ij}= |\phi_i,\phi_j\rangle\langle \phi_i, \phi_j |
$$
which leads to the fidelities
\begin{equation}
  \label{eq:fidelities}
  {\cal F}[\rho_{ij}^{\rm (dimer)},\rho_{ij}^{\rm (spiral)}]=\left\{
    \begin{array}{l r}
    \frac{\sin^{2S}(\frac{|\tilde{\phi}|}{2})}{(2 S+1)}   & \mbox{in the same rung}    \\
    \frac{1}{(2 S+1)}    & \mbox{otherwise} 
    \end{array}\leq \frac{1}{2}
\right. \,.
\end{equation}
These low values of fidelity indicate a very poor accuracy in the estimation of some mean values provided by MFA.
A typical way to improve these results consists in implementing a symmetry restoration over the mean field results\footnote{Here we assume that the true GS is
non-degenerate, and hence, is $SU(2)$ invariant, belonging to the $j_T=0$ sector.}:
\begin{equation}
  |\rm{SR\, MF}\rangle = \int {\bf R}_{\Omega} |{\rm MF}\rangle    d\mu_{\Omega}
\end{equation}
where $d\mu_{\Omega}$ is the normalized invariant measure of $SU(2)$ and ${\bf R}_{\Omega}$ is a given global rotation.   The state $|\rm{SR\, MF}\rangle$ is then
 the projection of the $|\rm MF\rangle$ over an eigenspace of the total angular momentum operator ${\bf J}_t^2$, built as a coherent superposition of all other states obtained by global rotations\cite{RS.80}.
 In finite size systems ($N,S<\infty$) a suitable choice of the representation of ${\rm R}_{\Omega}$ leads 
 to reduce the energy associated to $ |\rm{SR\, MF}\rangle$ slightly below the mean field result. 

The local state of a given subsystem ${\cal A}$ is obtained as the partial trace
 over the complementary subsystem $\bar{\cal A}$:
\begin{subequations}
\begin{eqnarray}
  \rho_{\cal A}&=&{\rm Tr}_{\cal A}|{\rm SRMF} \rangle\langle {\rm SRMF} |\\
&=&
\int\!\!\!\!\int  {\bf R}^{\cal A}_{\Omega} \rho_{\cal A}^{\rm spiral}  ({\bf R}^{\cal A}_{\Omega'})^\dagger
w_{\cal A}[\Omega,\Omega']d\mu_{\Omega}d\mu_{\Omega'} \label{eq:SRint1}\\
\rho_{\cal A}^{\rm spiral}&=&| {\rm spiral} \rangle\langle {\rm spiral} |_{\cal A}\\
w_{\cal A}[\Omega,\Omega']&=&\langle {\rm spiral}|({\bf R}^{\bar{\cal A}}_{\Omega'})^{\dagger} {\bf R}^{\bar{\cal A}}_{\Omega} |{\rm spiral}\rangle_{\bar{\cal A}} 
\label{eq:SRintw} 
\end{eqnarray}
\end{subequations}

Notice that the integrand in (\ref{eq:SRint1}) is not hermitian, but the integral is it. For large $S$ 
or $n_{\bar{\cal A}}$,  $w_{\cal A}[\Omega,\Omega'] \rightarrow \delta(\Omega-\Omega')$ (see \ref{sec:wA}) and hence\footnote{This result also  corresponds to average the reduced state over all possible mean field solutions. It can also be derived through a path integral approach as the Static Path Approximation\cite{CMR.07}. }
$$
\rho^{\rm SR}_{\cal A}\rightarrow \int {\bf R}^{\cal A}_{\Omega}
\rho^{\rm spiral}_{\cal A} ({\bf R}^{\cal A}_{\Omega})^{\dagger} d\mu_{\Omega}\,.
$$ 
Due that the energy depends on the pairs local states, it implies that the symmetrized state in the large $N$ limit has the same energy than the spiral one. 

Exploiting the structure of the state, we can rewrite the local statistical operator as $\rho^{\rm SR}_{\cal A}=\sum_j p^{\cal A}_j \Pi^{\cal A}_j$ where $\Pi^{\cal A}_j$ is the projector over the total spin in ${\cal A}$ and
$p_j^{\cal A}=\langle {\rm spiral}|\Pi_j^{\cal A} |{\rm spiral}\rangle$. 

\begin{figure}[t!]
\begin{center}
\includegraphics[width=0.8\columnwidth]{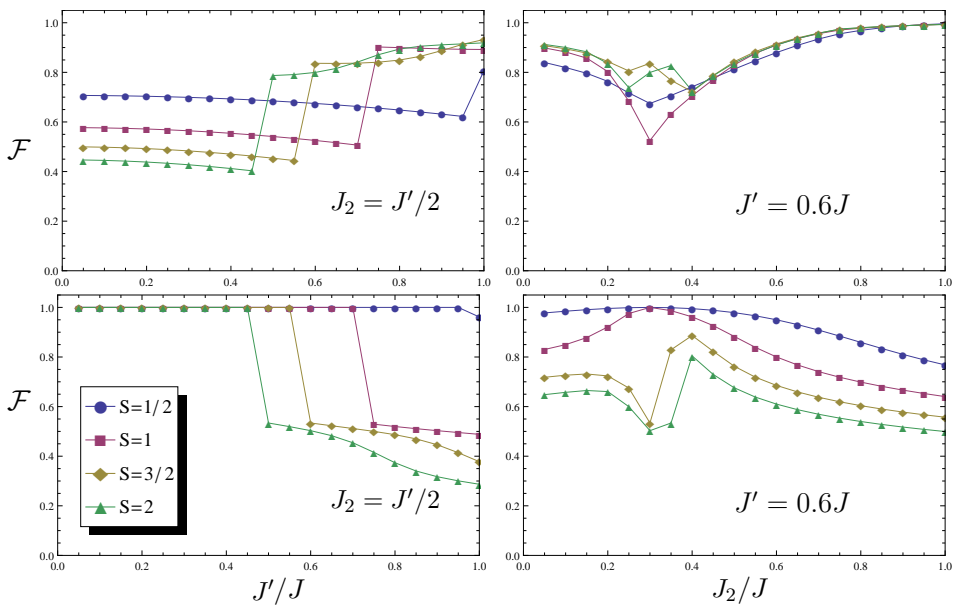} 
\end{center}
  \caption{Fidelities between the exact and  the approximate local strong coupled pairs states in the local mean
  field approximation + symmetry restoration (Top) and to the singlet state (Bottom) over the line $J'=2J_2$ (Left) and $J_2=0.6 J$ (Right) for different values of $S$. We can appreciate that for some $J_c^{'}$ the fidelity of the exact local state and the singlet state is drastically reduced. In the left panels, we can observe that
as for small $J'$ the local state match exactly with the singlet state, being this state poorly approximated by a product state. For larger $J'$,
the local state change sharply, becoming in the spiral state in a good approximation.
In the right panels, the behavior of  the fidelities with $J_2$ for $J'=0.6J$ is depicted. We can see that the dimerized region around the $J'=J_2$ line becomes sharper as the spin grows up, and then the local state moves away the dimerization for $S$ larger than the critical value. }
  \label{fig:fidelities}
\end{figure}

In order to check the accuracy of the approximation, we will consider the state of the subsystem defined by a pair of spins inside a same rung $\rho_{\rm rung}$.
 In the top panels of Figure \ref{fig:fidelities}, ${\cal F}[\rho_{\rm rung}^{ex},\rho_{\rm rung}^{\rm SRMF}]$ is depicted for different $S=1/2,1,3/2$ and $2$, over the line $J'=2J_2$, as a function of $J'$ (left) and over the line $J'=0.6J$ as a function of $J_2$ (right)\footnote{In this case, the reference values for the local density matrices were obtained by means of the Lanczos method for numerical exact diagonalization (see  \ref{sec:nummeth}).}. On the 
bottom panels, the corresponding fidelities to the local singlet state are depicted. On the one hand, we observe that over the dimerizing line $J'=J_2/2$ the 
fidelity between the exact and the SRMF state is reduced as the spin grows up, and is sightly decreasing, up to the point in which the $|{\rm dimer}\rangle$
ceases to be the GS. At this point, the fidelity gives a jump, and start to increase with $J'$. On the other hand, we observe that if we move 
crossing the dimerizing line with constant $J'/J$, the fidelity with respect on the SRMF local state has a minimum in those cases for which the system reach
dimerization (for the $J'=0.6 J$ case, $S=1/2,1$), whereas when the dimerization is not reached, (for $J'=0.6J$, $S=3/2,2$) the behavior is the opposite. At the same 
time, the fidelity with the singlet state falls of  up to a value of $0.5$, and continues decreasing with $J'$.  
\begin{figure}[t!]
\begin{center}
\includegraphics[width=0.7\columnwidth]{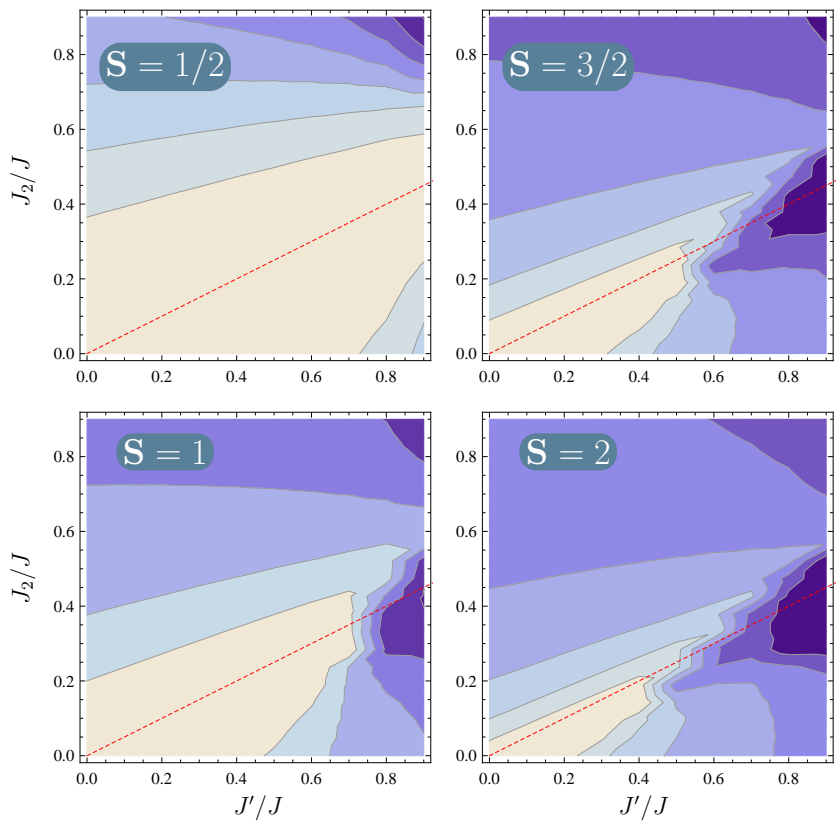} 
\end{center}
\caption{(Color Online)
Landscape of the fidelity of the exact local state against the singlet state for $S=1/2$ (Top,Left),
$S=1$ (Bottom,Left), $S=3/2$ (Top,Right), and $S=2$ (Bottom,Right). The dashed red line corresponds to $\frac{|2 J_2-J'|}{J}=0$. We can appreciate how the region where the 
local state is near to the singlet state is reduced as the spin grows up. Besides, we can observe that the width of the region around the line $\frac{|2 J_2-J'|}{J}=0$ where the dimerization exists is nearly constant (as is predicted by RPA) but becoming slightly more narrow as we approach to $J_c'$.
}
\label{fig:fidelities3}
\end{figure}
\subsection{Composite unit cell mean field approach for the $S=1/2$ case}
As we say before, the ``classical'' mean field is not suitable to represent the system in a dimerized phase. In particular, for $S=1/2$, we know that the 
exact solution corresponds to such a phase for $2 J_2-J'=0$ and $J'<J$. In this way, a better starting point to improve the mean field approximation is to choose as the unit cell those pairs of spins coupled by the stronger interaction. For instance, if $J>J_2,J'$,  we will look for mean field solutions of the form
\begin{equation}
  \left|{\rm MF}_{\rm pair}\right\rangle= \bigotimes_{i=1}^{N} |\alpha_{i}\rangle \label{mfss1b}
\end{equation}
where $|\alpha_i\rangle \in {\mathcal H}_{2\,i-1}\otimes {\mathcal H}_{2\,i}$. 
In order to simplify the further discussion, we introduce the parameter $\gamma=\frac{2J_2-J}{J}\frac{S(S+1)}{3/4}$ which, together with  $J'$, determines each point in the phase diagram. For $\gamma=0$, the lowest energy mean field state is obtained for  $|\alpha_i\rangle=|{\rm singlet}\rangle$, and coincides with the ``exact'' GS (\ref{eq:dimerstate}).
For small enough $|\gamma|$ we can expect that the true ground state results similar to the dimer state.  To drive the mean field optimization problem to a simpler form, we will restrict a little bit the family, but in a way that it still contains both the spiral ordered ``classical'' state as well the dimerized state.
 This can be achieved by setting $|\alpha_{i}\rangle=({\bf R}(\theta))^{i}|\alpha_{1}\rangle$ with ${\bf R}(\theta)$ a fixed planar rotation. 
 Also, due to the global rotational symmetry, we can choose $|\alpha_{1}\rangle$ as an state with
 total spin $\langle\vec{\bf J}\rangle$ polarized in the $x$ direction, and both local spin mean values laying on the $xz$ plane. With these assumptions, we can face in a analytical way the mean field problem for the $S=1/2$ case: in such a case, we have to consider a one-parameter family:

\begin{eqnarray}
|\alpha_1[\zeta,\phi,\tau]\rangle &=& \cos(\frac{\zeta}{2})( \sin(\phi/2)  |{\rm singlet}\rangle +  \cos(\phi/2) i  |y\rangle)- e^{i\tau}\sin(\frac{\zeta}{2})|z\rangle \label{eq:shtwomf}
\end{eqnarray}

where $|\mu\rangle$ ($\mu=x,y,z$) are the states such that ${\bf J}^2|\mu\rangle=2|\mu\rangle$ and ${\bf J}_\mu|\mu\rangle=0$, 
and $\zeta$, $\phi$, $\tau$ and $\theta$ are parameters ranging on $0\leq\zeta<\frac{\pi}{2}$, $0\leq\phi<\pi$, 
$-\pi/2\leq\tau\leq \pi/2$,$-\pi<\theta<\pi$. 

It is easy to verify that $\tau$ does not modifies the internal energy of the pair, but just the magnitude of the local spin mean values. For this reason, the optimum state always belongs to the subset $\tau=0$. The remaining parameters have a direct interpretation:
 the local mean field  spin values have equal magnitudes $|\langle \vec{\bf S}_{i}\rangle|=\frac{\sin(\zeta)}{2}$, being $\phi$ the angle between them. With this parametrization, 
 the separable case in Eq. (\ref{mfss1}) is obtained for $\zeta=\pi/2$  whereas for $\zeta=0$ the $|{\rm singlet}\rangle$ is recovered.  
For the general case, the variational energy is given by
\begin{eqnarray}
\nonumber
\frac{E_{{\rm MF_{\rm pair}}}[\theta,\zeta,\phi]}{J N/4}&=&
\cos (\phi )+\cos(\zeta)(\cos(\phi)-1)+\\ 
&&  \hspace{-1.5cm}+\sin^2(\zeta ) (\gamma \cos (\theta )  +\frac{2\,J'}{J} \cos(\theta-\frac{\phi}{2}) \cos(\frac{\phi}{2}))\,.
\label{mfpairenergysh}
\end{eqnarray}
 For $\zeta=\pi/2$, the previous expression is reduced to (\ref{mffenergy}), corresponding to a factorized
 state, but is easy to check that the minima  of (\ref{mffenergy}) does not correspond to global minimum for (\ref{mfpairenergysh}). 
  For $|\gamma|<1$, the true minimum is attained for $\zeta=0$, $\phi=\pi$, corresponding to the dimerized state
  (\ref{eq:dimerstate}), with energy $E_{{\rm MF_{pair}}}[\theta,\zeta,\phi]=E_{dimer}=-\frac{3}{4} J N$,
  which means that we can expect a dimerized phase in all this region. 
 
On the other hand, for  $|\gamma|>1$, but keeping $|J'|<J$ the minimum is attained for
$\phi=\pi$, $\cos(\theta)=\frac{-\gamma}{|\gamma|}$ and $\cos(\zeta)=\frac{1}{|\gamma|}$,
with energy $E_{{\rm MF_{pair}}}[\theta,\zeta,\phi]=-\frac{1}{4} J N (\frac{1}{|\gamma|}+1+|\gamma|)$.
 This phase is characterized by a breaking of the $SU(2)$ symmetry but, differently to the large $S$ prediction,
 this phase results colinear instead of spiral.
 Another difference is related with the degree of symmetry breakdown: for $|\gamma|\gtrsim 1$, the 
 difference between the the local states associated to
 the corresponding mean field state and such obtained by 
the symmetry projection method discussed before are not very important. On the other hand, as $|\gamma|$ grows, 
the mean field becomes near to the single site 
mean field result, for which the symmetry projection is required.
%
%
%
\subsection{Generalization to the Spin $S$ case and classical limit arising}

For  $S>1/2$, the fully analytical treatment discussed in the previous paragraph is not feasible anymore, because the local basis grows as $S^2$.
For small $S$  we can try to solve the full mean field equation numerically  by a self-consistent treatment or any other high dimension optimization method. 
 However, in order to look for analytical solutions, we can search it in a reasonable reduced family. Probably, a  good starting point is given by the family of
 states generated by the singlet state and a general fully factorized spiral state (\ref{eq:plspiralstates}). 
Up to a global rotation, the most general local state in this family is given by 
\begin{equation}
  |\psi_{\zeta,\tau,\phi,}\rangle={\cal N}(\cos(\zeta/2) |{\rm singlet}\rangle + \sin(\zeta/2) e^{i \tau}|\phi\rangle) 
\end{equation}
with ${\cal N}^{-1}=\sqrt{1-\frac{\sin(\zeta)\cos(\tau)\sin^{2S}(\phi/2)}{\sqrt{2 S+1}}}$. Indeed, the solution for the pair mean field
in the $S=1/2$ case always belongs to this family. The energy associated to this family of states is given by
\begin{eqnarray}
\nonumber \frac{  E_{{\rm MF_{pair}}}[\theta,\phi,\zeta,\tau]}{{\cal N}^{2}}&=&E_{dimer} \; (\cos^2(\zeta/2)+\sin(\zeta)({\cal N}^2-1))+\\
&& + \sin^2(\zeta/2) E_{\rm sep}[\theta,\phi]
\end{eqnarray}
being $E_{\rm ESP}[\theta,\phi]$ the energy associated to the fully separable spiral state defined in (\ref{mffenergy}). 
For $S\leq 3/2$, over the line $J'=2 J_2$ the
 energy is always optimized  for the fully-dimerized state, while for $S\geq2$, the optimum value is attained at the
 fully factorized classical state for $J'>J'_c=2 J/S$, what can be see as a first order phase transition. This result could sound disappointing: a direct mean field treatment, even if we consider the complex
 cell proposed here, is not able to reproduce the transition between the dimerized and the spiral phases. However,
examining the behavior of the fidelities between the exact state and the single site mean field + symmetry restoration  (see Fig \ref{fig:fidelities}), we observe that at least locally, after the transition the local state of the strong correlated pairs tends to the mean field result, including the cases for which the pair mean field solution predicts dimerization. In this way, we can not expect to obtain
a better mean field solution by exploring a larger family, leading to understand it as a genuine collective
full quantum effect, coming from the quantum corrections to the correlation energy on the excited states.
This point will be discussed  more detailedly in the next section.

\section{Low-lying excitations around the dimerized phase and gap estimation}
\label{sec:gap}

In the previous sections, we have discussed the structure of the GS of the system for both $S=1/2$ and $S\gg 1$ near the line $0<J'=2 J_2<J$. Then, we saw that for $S=1/2$ the system is exactly dimerized, evolving to a N\'eel ordered state as we move away this line while, for the large $S$ case, the system presents spiral order.
 In this section we will complete this picture analyzing the low-lying elementary  excitations of this system, as well the behavior of its pair correlations. Also, we are interested into determine in which way quantum fluctuations corrects the energy of the excited states from the mean field result,  leading to the crossover observed through the exact diagonalization. 

Any of the previous mean field approximations to the GS are always limited by the the lack of quantum 
correlation in the variational ansatz among the  different components, which was essential to make the problem tractable. 
However, always looking for analytical results, we can improve the approximation in many ways.  
If what we are looking for is just a better approximation to the GS energy, a perturbative expansion on the couplings 
could be feasible.
 On the other hand, if we want to look for the energy spectrum or estimate correlations, perturbative theory becomes cumbersome. 
 A more handsome way to get all at the same time is through the Random Phase Approximation (RPA)\cite{RS.80,MRC.10}. 
 When is applied to obtain the GS, this approximation consists on a generic prescription to build a ``Gaussian'' 
approximation  to the GS\cite{MRC.10} starting from a mean field state, which implies a certain kind of 
approximated bosonic mapping.
Indeed, it corresponds to the approximate local bosonic map:
\begin{subequations}
  \label{eq:rpabosscheme}
\begin{eqnarray}
  |0\rangle &\rightarrow& |0\rangle_{i,{\rm bos}}  \\
  |\alpha\rangle\langle 0|_{i} &\rightarrow &{\bf a}_{\alpha,i}^{\dagger}\\
  |\alpha\rangle\langle \alpha|_{i} &\rightarrow &{\bf a}_{\alpha,i}^{\dagger}{\bf a}_{\alpha,i}\\
  |0\rangle\langle 0 | &\rightarrow& 1-\sum_{\alpha} {\bf a}_{\alpha,i}^{\dagger}{\bf a}_{\alpha,i}
\end{eqnarray}
\end{subequations}
being $|0\rangle_i$ the GS of the local mean field Hamiltonian and $|\alpha\rangle$ its excited states.
It is straightforward to check that this map
 preserves the mean values of any commutator of two operators over the mean field state.
 In the next paragraphs we will consider the RPA treatment built over the single site and the two sites  mean field
 state. Finally, we analyze the correction to the gap and the critical point localization 
 over the line $\gamma=0$  through a perturbative treatment.
\subsection{Random phase approximation over the spiral mean field solution}
If we start from a single site mean field, RPA scheme (\ref{eq:rpabosscheme})  always leads to the local approximate bosonization 
\begin{subequations}
  \label{eq:RPAss1}
\begin{eqnarray}
{\bf S}^{x'}_{i} &\rightarrow &\sqrt{2 S}\frac{{\bf b}_{i}+{\bf b}_{i}^\dagger}{2}  \\
{\bf S}^{y'}_{i} &\rightarrow& \sqrt{2 S}\frac{{\bf b}_{i}-{\bf b}_{i}^\dagger}{2 {\bf i}}\\
{\bf S}^{z'}_{i} &\rightarrow&  S (1-{\bf b}_{i}^\dagger{\bf b}_{i}/S)
\end{eqnarray}
\end{subequations}
where $\mathbf{S}^{\mu'}_{i}$ are the components of \emph{local} spin operator in an \emph{intrinsic} basis $(x',y,z')$ 
chosen in a way that the local spin polarization points in the $z'$ direction. 
This bosonization leads to a spin wave-like Hamiltonian
$$
{\bf H}_{\rm SW}=E_{\rm sep}-\frac{{\rm Tr}\Lambda}{2} + \frac{1}{2}\left({\bf z}\,{\bf z}^\dagger \right){\cal H}\left(^{\bf z}_{{\bf z}^\dagger} \right)
$$
with ${\bf z}={{\bf b}_1,\ldots {\bf b}_{2N}}$ a ``vector'' of bosonic operators,  $\Lambda={\rm diag}(\varepsilon_{i}^{(\rm MF)})$ a diagonal matrix with elements equal to the local mean field Hamiltonian excitation energies, and   ${\cal H}$ a matrix with the block form
$${\cal H}=\left( ^{\Lambda+\Delta^{+}}_{\;\;\overline{\Delta^{-}}} \; _{\Lambda+\overline{\Delta^{+}}}^{\;\;\Delta^{-}}\right)\,.$$
For the translational invariant coplanar mean field case (Eq. \ref{eq:plspiralstates}), we can diagonalize the quadratic form analytically: starting with the Fourier transform,
$$
\tilde{\bf b}_{k,j}=
\frac{1}{\sqrt{N}}
\sum_{i=1}^{N} 
e^{-\frac{\bf{i} 2\pi}{N} k i }
{\bf b}_{2 i + j} 
\hspace{1cm} 
^{j\in \{0,1\} }_{k \in\{0,\ldots, N-1\}}
$$
we can rewrite the Hamiltonian as
$$
{\bf H}=E_{\rm sep}- N \lambda + \frac{1}{2}\sum_{k=0}^{N-1} \left(\tilde{\bf z}_{k} \,\tilde{\bf z}_{N-1-k}^\dagger \right){\cal H}_k\left(^{\tilde{\bf z}_k}_{\tilde{\bf z}_{N-1-k}^\dagger} \right)
$$
where 
$$\lambda=\varepsilon_{i}=S |J \cos(\phi)+J' \cos(\theta-\phi)+2 J_2 \cos(\theta)) |$$ and
\begin{equation}
  {\cal H}_{k}=\left(\begin{array}{cc} 
\lambda +\Delta_k^+ &  \Delta_k^-\\
\Delta_{k}^- &  \lambda+ \Delta_{k}^+
\end{array}
\right)
\end{equation}
with $\Delta_k^{\pm}$ blocks defined by

\begin{eqnarray*}
  \Delta_{k}^+&=&
  \left(\begin{array}{cc}
2 J_2  \cos(k) \cos^2\left(\theta/2\right)   & J\cos^2\left(\phi/2\right) +J' e^{-i k} \cos^2\left(\frac{\theta-\phi}{2}\right)\\
J\cos^2\left(\phi/2\right) +J' e^{i k} \cos^2\left(\frac{\theta-\phi}{2}\right) &  2 J_2  \cos(k) \cos^2\left(\theta/2\right)
\end{array}\right)S  \\
  \Delta_{k}^-&=&-
\left(\begin{array}{cc} 
2 J_2  \cos(k) \sin^2\left(\theta/2\right)   & J\sin^2\left(\phi/2\right) +J' e^{-i k} \sin^2\left(\frac{\theta-\phi}{2}\right)\\
J\sin^2\left(\phi/2\right) +J' e^{i k} \sin^2\left(\frac{\theta-\phi}{2}\right) &  2 J_2  \cos(k) \sin^2\left(\theta/2\right)
\end{array}\right)S 
\end{eqnarray*}
Now,  by means a canonical transformation we diagonalize each ${\cal H}_k$. We can recover the excitation energies in terms of invariants of the ${\cal H}_k$ matrices:
$$
\omega_{\pm}(k)=\sqrt{\frac{{\rm tr} ({\cal M}{\cal H}_k)^2 \pm\sqrt{({\rm tr} ({\cal M}{\cal H}_k)^2)^2-16 \det{\cal H}_k} }{2}}
$$
Zero modes arise when $\det{\cal H}_k=0$. This happens always for $k=0$ and $k=\tilde{\theta}$, corresponding to 
Goldstone modes associated global rotations around the global $y$ axes or an axes over the global  $x y$  plane respectively.

\subsection{Random phase approximation over the fully-dimerized pair mean field}
Around the line $\gamma=0$, we know that for $S=1/2$ the mean field state is always the dimerized state (\ref{eq:dimerstate}), and that,
for small enough $J'$, there is always a region for which this result holds for any spin $S$. To obtain the RPA Hamiltonian associated to this mean field solution is useful to start from the expression (\ref{hjk}).
 Due to the structure of the Hamiltonian, the relevant local basis for this problem is given by $(|{\rm singlet}\rangle,|x\rangle,|y\rangle,|z\rangle)$ where $|\mu\rangle=\sqrt{\frac{3/4}{S(S+1)}}{\bf K}_{\mu}|{\rm singlet}\rangle$.
 Using the algebraic properties of the ${\bf K}_{\mu}$ operators defined in (\ref{eq:KopDef}) we can check that this is an orthonormalized basis. Following the RPA prescription (Eq. (\ref{eq:rpabosscheme})) we obtain the approximate  bosonic map:
 \begin{subequations}
   \label{eq:dimerbosonization}
\begin{eqnarray}
  \label{eq:bosdimergs}
|0\rangle &\rightarrow& |{\rm singlet}\rangle\\  
{\bf K}_{i,\mu}&\rightarrow& \sqrt{\frac{S(S+1)}{3/4}}({\bf a}_{i,\mu}+{\bf a}_{i,\mu}^\dagger)\\
{\bf J}_{i,\mu}&\rightarrow& {\bf 0}\\
{\bf J}^2_{i,\mu}&\rightarrow&  2\,{\bf a}_{i,\mu}^\dagger{\bf a}_{i,\mu}\,.
\end{eqnarray}
\end{subequations}
It leads to the bosonized Hamiltonian
$$
{\bf H}\rightarrow E_{dimer}+ \sum_{\mu} {\bf h}^{\rm bos}_{\mu}
$$
with 
\begin{eqnarray}
\nonumber
{\bf h}_{\mu}^{\rm bos}&=&  J \sum_i\left[  {\bf a}_{\mu,\,i}^\dagger{\bf a}_{\mu,\,i}+\right.\\
&&\hspace{-.6cm}\left.+\frac{\gamma}{4} ( {\bf a}_{\mu,\,i}^\dagger+{\bf a}_{\mu,\,i})( {\bf a}_{\mu,\,i+1}^\dagger+{\bf a}_{\mu,\,i+1})\right]
\end{eqnarray}

This is, in this approximation the system is mapped to three decoupled bosonic systems with first neighbor quadratic interactions. For $\gamma=0$, the systems become decoupled, due to the fact that the mean field state corresponds to a true GS. For general $\gamma$, the excitation energies of any of ${\bf h}_{\mu}^{\rm bos}$ are given by
\begin{equation}
  \label{eq:wRPAdim}
  \omega_{k}=J \sqrt{1-\gamma\cos\left(\frac{2\pi}{N}k \right)} .
\end{equation}
The correction to the ground state energy is hence given by
\begin{equation}
\label{eq:DEnRPADimer}
\frac{\Delta E_{{\rm RPA},{\rm dimer}}}{E_{dimer}}=2
\left(1-\frac{ 
\sqrt{1+\gamma}{\rm E}(\frac{2}{1+\gamma^{-1}})+
\sqrt{1-\gamma}{\rm E}(\frac{2}{1-\gamma^{-1}})}
{\pi}\right)
\end{equation}
where ${\rm E}(x)=\int_0^{\pi/2} \sqrt{1-x \sin^2(u)}du$  is the Elliptic Integral of second kind.

\begin{figure}[h]
\begin{center}
\scalebox{2}{\includegraphics{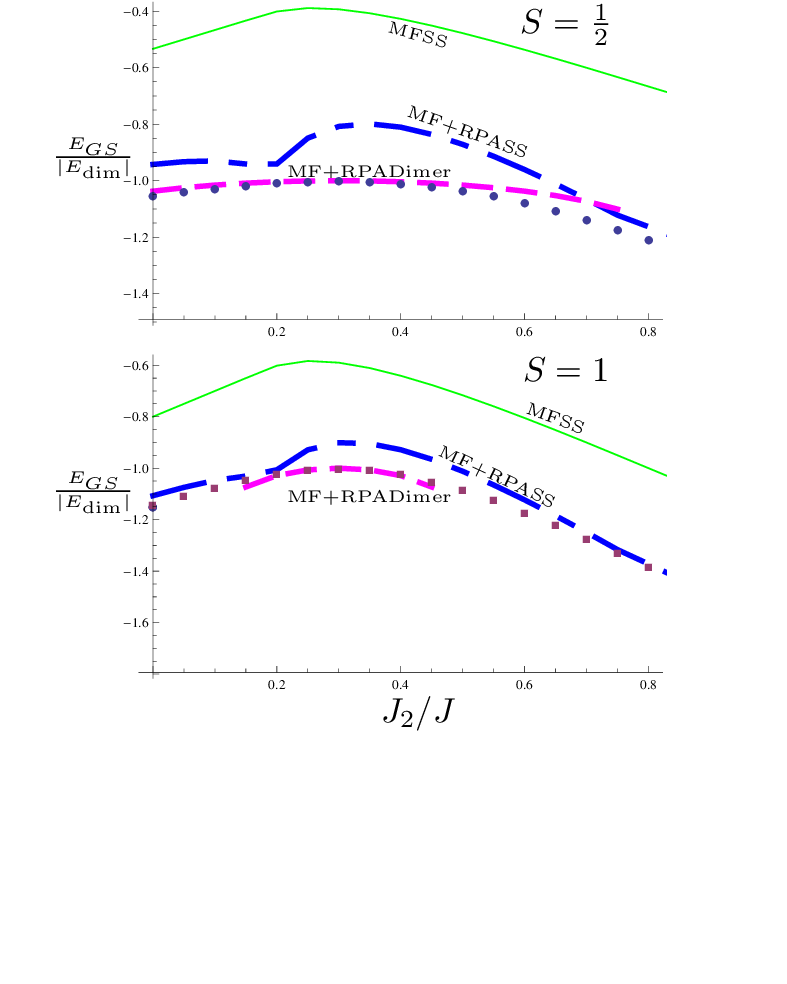}}
  \caption{(Color Online) Exact, mean field and RPA results for the ratio $E/|E_{\rm dimer}|$ over the line $J'=0.6 J$ for a spin $S=1/2$ (left) and $S=1$ (right) over a $2\,N=40$ spin chain. Isolated dots corresponds to the exact result.}
  \label{fig:rpares}
\end{center}
\end{figure}
In Figure \ref{fig:rpares} the exact (dots), mean field (solid line)  and the RPA results (dashed lines) based on simple and double unit cells for the $E/E_{\rm dimer}$ ratio over the line $J'=0.6J$ are depicted for the spin $S=1/2$ and $S=1$ cases, over a $2\,N=40$ sites periodic chain. We can appreciate that the RPA correction over the single site mean field (blue dashed line) improves significativelly the bare mean field result, but it is not able to give an accurate description of its behavior over the dimerized region. However, through the double unit cell MF+RPA approach (purple dashed lines) we found a very accurate description which matches the exact result over the point $J_2=J'/2=0.3J$. Outside of the region where the fully-dimerized mean field is stable, the single site approach becomes closer to the exact result, in agreement with the observed behavior of the fidelity between the exact ground state and the mean field state (\ref{fig:fidelities}).
  
An interesting point is that the  bosonized Hamiltonian is stable if $|\gamma|< 1$.
It give us a different way to determine the region where the dimerized state is stable when we move in the parameter space keeping $J'/J$ constant:  as the local spin grows up, the width around the $\gamma=0$ line where the system presents dimerization is reduced as $\approx 1/S^2$. This estimation is approximately what we can appreciate in the Fig  \ref{fig:fidelities3}.  
However, from the same figure we can also see that the prediction of a width independent of $J'/J$ is not completely fulfilled: this difference seems to be related with the inability of this method to detect the crossing level over the $\gamma=0$ line.

\subsection{Higher order corrections and gap estimation over the $\gamma=0$ line}
 As we say in the previous paragraph, RPA is not able to detect correctly the phase transition due the fact that for $\gamma=0$ all corrections vanish, predicting a constant finite gap $\Delta\varepsilon=J$.
 An elementary way to understand the behavior of the gap and the structure of the excited state consists in develop it as a perturbative expansion. A practical way to do it is again to start from the expression (\ref{hjk}) for the Hamiltonian around $J'=J_2=0$.
 Due the large degeneration on each energy subspace, the first step for a perturbative expansion is to choose a suitable basis for each subspace.
 To do this, we begin by observing that both unperturbed and full Hamiltonian commutes with the total angular 
 momentum operator ${\bf J}_{T}^2= \sum_{i,j} \vec{{\bf S}}_{i}\cdot\vec{{\bf S}}_{j}- 2 N S (S+1)$, so we can split the problem on 
 each total momentum sector.
 A second observation comes form the fact that  the spectrum on each site is controlled by the local total momentum ${\bf J}_i^2$. As we are looking for the behavior of the GS, we can limit to consider state with global total
 momentum $j_T=0$ or $j_T=1$, due that sectors with larger $j_T$ involve states with unperturbed energies
 three times bigger than the gap associated to the unperturbed Hamiltonian.
 It allows us to consider just local excitations on the $j_i=1$ sector. Finally, another symmetry in the Hamiltonian is related with the translational invariance. A suitable basis for the relevant subspace is hence
 given by ${\cal B}=\bigcup_{j,\varepsilon,k}{\cal B}_{j,\varepsilon,k}$ where $\varepsilon$ is the (unperturbed) energy associated to a given sector (relative to $E_{\rm dimer}$). An explicit expression for ${\cal B}_{j,\varepsilon,k}$ is given by
\begin{equation}
\begin{small}\begin{array}{|c | c | c | c | c| c}
\hline
{\cal B}_{j,\varepsilon,k} & \varepsilon=0 & \varepsilon=J & \varepsilon=2J & \varepsilon=3J & \ldots\\
\hline
j=0           & \{|{\rm dimer}\rangle\} & - & \{|j,k\rangle \} & \{|lm,k\rangle\}&\ldots\\
j=1           & - & \{ |\mu,k\rangle \} &  \{ |j\eta,k\rangle \} & \ldots & \ldots\\
\ldots        & - &  - &\ldots &\ldots  & \ldots
\end{array}
\end{small}
\end{equation}
with
%
\begin{subequations}
\begin{eqnarray}
  \label{eq:s1states}
|\mu,k\rangle&=& \sum_i\frac{e^{-{\bf i} i k }}{\sqrt{N} {\cal N}}{\bf K}_{\mu,i}|{\rm dimer}\rangle  \\
|j,k\rangle&=& \sum_i\frac{e^{-{\bf i} i k }}{\sqrt{3 N} {\cal N}^{2}}{\bf K}_{i}\cdot{\bf K}_{i+j}|{\rm dimer}\rangle\\
|j\eta,k\rangle&=& 
           \sum_{i,\mu,\nu}\frac{e^{-{\bf i} i k }\epsilon_{\mu\nu\eta}}{\sqrt{2 N}  {\cal N}^{2}}{\bf K}_{\mu,i}{\bf K}_{\nu,i+j}|{\rm dimer}\rangle\\
|lm,k\rangle&=&\!\!\!\!
              \sum_{i,\mu\nu\eta} \!\!\frac{e^{-{\bf i} i k }\epsilon_{\mu\nu\eta}}{\sqrt{6 N} {\cal N}^{3}}{\bf K}_{\mu,i-l}{\bf K}_{\nu,i}{\bf K}_{\eta,i+m}
|{\rm dimer}\rangle
\end{eqnarray}
\end{subequations}
being  ${\cal N}=\sqrt{\frac{S(S+1)}{3/4}}$, $1\leq j\leq N/2$, $1\leq m,l\leq N/3$, and $\mu,\nu,\eta=x,y,z$. 
Before giving the expressions for the corrections to the spectrum, we are going to give an interpretation of the interaction terms in (\ref{hjk}) in terms of its action on each sector.
On the one hand, the quadratic terms in ${\bf J}_{\mu,i}$ acts over states with excitations on contiguous sites $i$ ``rotating'' both local excitations, and preserving the number of excitations (i.e., connecting states just on the same sector).
 Because it annihilates pairs of contiguous singlets, it has no effect on the  sector with $\varepsilon=J$ and on any state with excitations on non-contiguous sites.
 In this way, this term is diagonalized \emph{exactly} on each subspace ${\cal B}_{j,\varepsilon}$ by the basis  given in (\ref{eq:s1states}). On the other hand, the bilinear terms in  ${\bf J}_{\mu,i}$ and ${\bf K}_{\nu,j}$ creates (or destroys) excitations over empty sites, depending on the occupation of its neighbor sites, and hence, connecting contiguous sectors with the same $j_T$.
 Because it does not  connect states inside the same sector, all its contribution shows at second order.

Defined the basis, we are ready to evaluate the corrections on each sector. Of course, the sector ${\cal B}_{0,0}$ corresponds to an eigenspace on the
full $\gamma=0$ line, with fixed energy $E_{\rm dimer}$.
 Now we will see the effect on the other sectors. In particular, we will look for the new GS on the sectors $j_T=0$ and $j_T=1$, considering unperturbed states with up to two excitations, i.e on the subspace generated by ${\cal B}_{1,J}$ and ${\cal B}_{0,2J}$.

\emph{${\cal B}_{0,2,k}$} On this subspace, both interaction terms in (\ref{hjk}) contributes to correct the energy. To analyze the corrections to the spectrum in this subspace, we will consider three different cases: $|1,k\rangle$,  $|2,k\rangle$,  and $|j>2,k\rangle$. The energy (relative to $E_{\rm dimer}$) 
associated to states with two contiguous excitations evolves as
\hspace*{-2.5cm}
\begin{subequations}
\begin{eqnarray}
 \frac{\Delta E_{|1,k\rangle}}{J}&=& 2-\frac{J'}{J} - 
\frac{S(S+1)}{3/4}\frac{\cos^2(k/2)}{2}\left(\!\!\frac{J'}{J}\!\!\right)^2\\
\frac{\Delta E_{|2,k\rangle}}{J}&=&  2-\frac{S(S+1)}{3/4}\frac{1}{4}\left(\!\!\frac{J'}{J}\!\!\right)^2\\
\frac{\Delta E_{|j>2,k\rangle}}{J}&=&2-\frac{S(S+1)}{3/4}\frac{1}{2}\left(\!\!\frac{J'}{J}\!\!\right)^2 \hspace{.2cm}
\end{eqnarray}
\end{subequations}

\begin{figure}[t!]
  \centering
\rput{90}(-0.25,4.){$J_c^{'}/J$}
\rput{0}(7.0,0){$S$}
  \includegraphics[width=.7\textwidth]{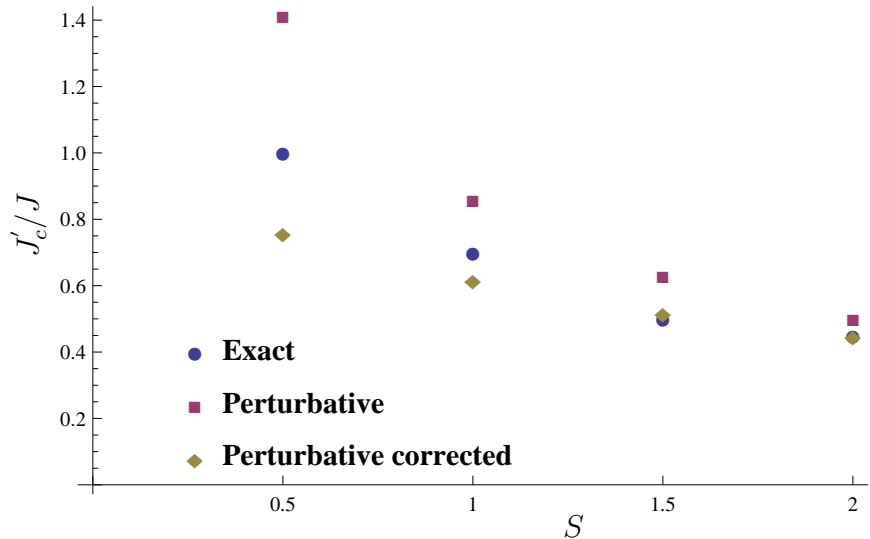}
  \caption{(Color Online) Estimation of the critical value of $J'/J$ over the line $J'=2J_2$ as a function of $S$. 
The \emph{exact} points were obtained by Lanczos diagonalization for a lattice with $N=6$ rungs. Largest deviations are observed for $S=1/2$, for which the exact critical point belongs to a non-perturbative regime. As the local spin $S$ grows, the critical value of $J'$ moves inside the perturbative regime. }
  \label{fig:critical}
\end{figure}

\emph{${\cal B}_{1,1,k}$}. This sector corresponds to states with a single excitation. As the quadratic term in ${\bf J}_i$ vanishes, at leader order the effect of the interaction is to mix this state with states in the ${\cal B}_{1,2,k}$ sector. At second order,
the energy of the corresponding state is given by
$$
 \frac{\Delta E_{|\mu,k\rangle}}{J}= 1-\frac{S(S+1)}{3/4} \left(\frac{J'}{J}\right)^2 \frac{\cos^2(k/2)}{2}
$$ 
which results three times degenerated, as we can expect from the $SU(2)$ symmetry. 
At second order, the GS on both ${\cal B}_{0,2,k}$ and ${\cal B}_{1,1,k}$ sectors becomes equal at $J'=J$, regardless $S$. However, at $J'=J\sqrt{\frac{3/2}{S(S+1)}}$ the GS on the  $j_T=1$ sector reach $E_{\rm dimer}$, becoming (on this approximation) the GS of
the full system.
  However, this is not the end of the story: numerical results show that the true GS after the phase transition does not belong to the $j_T=1$ but to the $j_T=0$ sector.
 For $S\geq 1$ we can reproduce this result and improve the approximation to the phase transition point by taking into account the ${\bf J}_i{\bf J}_{i+1}$ interaction terms \emph{exactly}. It leads to the following expressions for the GS on each sector:
\begin{subequations}
\begin{eqnarray}
\frac{\Delta E_{|1,0\rangle}}{J}&=&2 -\frac{J'}{J}- \frac{S(S+1)}{3/4}\frac{1/2}{1-\frac{J'}{J}}\left(\frac{J'}{J}\right)^2
\\
\frac{\Delta E_{|\mu,0\rangle}}{J}&=&1 -\frac{S(S+1)}{3/4}\frac{1}{2 (1-\frac{J'}{2 J})}\left(\frac{J'}{J}\right)^2
\end{eqnarray}
\end{subequations}
According to this result, the transition is reached at 

\begin{eqnarray}
J^{'}_{o}&=&J 
\frac{\sqrt{1+4 \frac{S(S+1)}{3/4}}-3}{\frac{S(S+1)}{3/4}-2}  \\
&\approx& \frac{J}{2}\sqrt{\frac{3/4}{S(S+1)}-2} +{\cal O}^2(1/S)
\end{eqnarray}
on the $J_T=0$ sector. Although for $S=1/2$ this expression predicts an non-existent transition at $J'=0.76 J$, for $S\geq 1$ gives
a reasonable approximation (see Fig \ref{fig:critical}).

The fail for the $S=1/2$ case can be attributed to the fact that at $J'/J\approx 0.8$, higher order perturbative corrections competes with the correction in the
 denominator. For larger $S$, the transition point moves away to a region for which the expansion results accurate. As regard the true structure of the ground
 state, we should note that in the neighborhood of the transition point, any state with approximately localized but ``diluted'' pair excitations coupled at $j_{\rm exc}=0$ gives a similar value for the energy, reducing the possibility to obtain an accurate description of the GS by this way. This feature also explains why MFA+RPA fails to reproduce the structure of the GS around this point: as we approach to the critical point, quantum correlations between rungs are
 no longer weak, and this approximation ceases to be valid \cite{MRC.10}.

\section{Summary and discussion}
\label{sec:conclusions}

In the present paper, the general $SU(2)$ invariant quantum spin-S Heisenberg model on the zig-zag ladder was investigated. A sufficient condition for the existence of a
fully-dimerized exact eigenstate was demonstrated for a wide
 subfamily of such systems and, for the translational invariant case, a sufficient condition for this fully-dimerized
 eigenstate be the true ground state was established.
 Besides, by means of a combination of numerical and analytical techniques, the existence of this phase for a general  value of the local spin was proven, showing that the region in the parameter
 space corresponding to the dimerized phase is reduced as the  magnitude of the local spin grows. 
 
 In this regard, unlike the typical picture of the classical limit as a gradual reduction of the width of the quantum fluctuations, here
 it arises as a sudden reduction of those fluctuations, related to a level crossing between two structurally different states,  the  dimerized
 (full-quantum) and a spiral (semi-classical) states.

Analytical results was achieved by a generalization of the variational Mean Field Approximation 
consisting in enlarging the size of the unit cell and study by means a full quantum treatment.
 Through the RPA formalism, we was able to describe small changes on the gap around the fully-dimerized line. However, this treatment was not able to detect changes on the gap along the factorizing line.
 The explanation of why it is in this way arises from a perturbative analysis: there is a region in which the GS consists on excitations with non Gaussian correlations.
  A possible way to overcome this problem consists on generalize the perturbative treatment over the RPA approximated GS, technique that we are currently on development. 

We have postponed the study of  multipartite entanglement, time evolution of the system and energy excitations for a forthcoming work. In particular, the study of excited states in the general case where  different couplings distribution can be studied is a very interesting topic that deserves a careful study.

\section*{Acknowledgments}
The authors thank to Prof. Raul Rossignoli and Prof. Daniel Cabra for their useful suggestions and discussions.
J. M. Matera is supported by CONICET and C. A. Lamas is supported by CONICET (PIP 1691) and ANPCyT (PICT 2013-0009).

\appendix
\section{Numerical methods}
\label{sec:nummeth}

For low values of $S$, suitable numerical methods have been developed in order to find the GS and its correlations. If also $2 N\log(2 S + 1)$ 
can be considered small enough, we can evaluate both the GS and excited states and its energies through Arnoldi-like techniques for sparse linear 
systems  \cite{GOLUB.96}, which allow us to recover all the relevant observables of the system. On the other hand, because the size of the eigenvalue problem grows as  $\exp(2 N\log(2 S + 1))$, we are in hard troubles if we want to analyze by this method the large $N$ and large $S$ behavior of the system. 

On the other hand, if we are interested just on the structure of the GS, methods bases on Density Matrix Renormalization Group (DMRG) provide an efficient way to explore the large $N$ limit. However, the performance of this method is spoiled when we are working with systems near a critical point, especially if we consider a not too small $S$ case. This is due to the fact that DMRG is based on the assumption that the local subsystems are not too entangled among them in its exact GS, in order to obtain an efficient and accurate description in terms of an small number of states, by looking for a suitable set of local state basis\cite{RO.97,Sch.05,Sch.11}. 
 In particular, this is true for spin $1/2$ (1-D) chains with local or quasi-local interactions if we are not too close to a critical point.  Another limitation of this technique is related with the fact that excited states do not satisfy the hypothesis of small inter-site entanglement, which limits the possibility to study quantities like the gap and the structure of the excited states with this technique.

In the present work, we use the Lanczos method as the reference method. However, in those cases when we need larger chains we take advantage of the DMRG method.  In both cases, we use the ALPS platform \cite{alpscollaboration}. Also, when we need to access to different local states (for instance, in order to evaluate fidelities), we have used a patch developed in our group for the ALPS code
 \footnote{The patch can be downloaded from:\\
 {\href{http://mauricio-matera.blogspot.com.ar/2014/03/information-theory-patch-for-alps.html}
 {http://mauricio-matera.blogspot.com.ar/2014/03/information-theory-patch-for-alps.html}}}
 allowing to access to this functionality.

\section{Properties of the Fidelity }
\label{sec:propfidelity}
The Fidelity is a measure of closeness between quantum states, defined as
\begin{equation}
  {\cal F}[\rho,\rho']={\rm Tr} \sqrt{\rho^{1/2} \rho' \rho^{1/2}}
\end{equation}
which for pure states is reduced to the absolute value of its overlap:
\begin{equation}
{\cal F}[|\alpha\rangle,|\beta\rangle]=\|\langle \alpha|\beta\rangle]\|  
\end{equation}
This measure is related to the Bures Angle
\begin{equation}
{ \Phi}[\rho,\rho']=\arccos({\cal F}[\rho,\rho'])  
\end{equation}
which defines a metric structure over the set of statistical operators. 

If two states have a fidelity near to 1, any observable evaluated on any of these two states must give similar
expectation values.
On the other hand, for states with low fidelity there is always a projective measure which can distinguish
between them with a high probability of success. 

We can relate the fidelity of the reduced local states to the global ones by
the inequalities
\begin{eqnarray}
{\cal F}[|\alpha\rangle,|\beta\rangle]&\leq&  {\cal F}[|\alpha\rangle_{\cal A},|\beta\rangle_{\cal A}]  \\
{\cal F}[|\alpha\rangle,|\beta\rangle]&\leq&  \sqrt{{\cal F}[|\alpha\rangle_{\cal A},|\beta\rangle_{\cal A}] 
{\cal F}[|\alpha\rangle_{\bar{\cal A}},|\beta\rangle_{\bar{\cal A}}]}
\end{eqnarray}
It means that despite a pair of global states would have a small fidelity, they can represent similar subsystem states. For instance, in certain contexts the mean field state can reproduce with good accuracy the behavior of local observables but could fail when we look for pair correlations.

\section{Evaluation of $w_{\cal A}[\Omega,\Omega']$}
\label{sec:wA}
To evaluate explicitly $w_{\cal A}[\Omega,\Omega']$ for the planar spiral case, we will observe first that  
$$
w_{\cal A}[\Omega,\Omega']=\prod_{i\in \bar{\cal A}}
\langle \varphi_i|{\bf R}_{\alpha \vec{n}} |\varphi_i\rangle
$$
where ${\bf R}_{\alpha \vec{n}} ={\bf R}_{\Omega'}^{\dagger} {\bf R}_{\Omega}$ is the composition of both rotations, which results in a rotation of an angle $\alpha$ around the $\vec{n}$ direction. Now, if the $|\varphi_i\rangle$ are polarized on the $\vec{n}_{\varphi_i}$ direction,  we can evaluate explicitly ea
ch $\langle\varphi_i|{\bf R}_{\alpha \vec{n}} |\varphi_i\rangle$ as 
$$\langle\varphi_i|{\bf R}_{\alpha \vec{n}} |\varphi_i\rangle=\left(\cos(\alpha)-{\bf i}\, \vec{n}\cdot\vec{n}_{\varphi_i}\sin(\alpha)  \right)^{2S}$$
Now we have to consider two different cases. On the one hand, if we suppose that $|\rm MF\rangle$ is an spiral state,  as $\vec{n}$ is a fixed versor but $\vec{n}_{\varphi_i}$ is 
changing, for almost all $i$ on a \emph{contiguous} block, $\vec{n}\cdot\vec{n}_{\varphi_i}<1$. In this way, $|\langle\varphi_i|{\bf R}_{\alpha \vec{n}} |\varphi_i\rangle|=1={\rm const}$ if and only if $\alpha=0$. For small $\alpha$ we get,
$$
w_{\cal A}[\Omega,\Omega']\approx \cos^{2 S N_{\bar{\cal A}}}(\alpha)\exp(- {\bf i} 2 S N \langle \vec{n}\cdot\vec{n}_{\varphi_i} \rangle_{\bar{\cal A}})
$$
where $N_{\bar{\cal A}}$ is the number of sites in $\bar{\cal A}$ and $\langle\vec{n}\cdot\vec{n}_{\varphi_i} \rangle_{\bar{\cal A}}$ is the mean value of the
 products over the sites in $\bar{\cal A}$. This quantity  vanishes if $\langle\vec{n}\cdot\vec{n}_{\varphi_i} \rangle_{\bar{\cal A}}$ is a large set of contiguous sites.

On the other hand, if the state on the subsystem $\bar{\cal A}$ is near enough to a N\'eel state, there is another possibility: because we can choose $\vec{n}$
 in a way that $|\vec{n}\cdot\vec{n}_{\varphi_i}|=1$, the family of rotations such that ${\bf R}_{\alpha \vec{n}}$ corresponds to a rotation around the magnetization axes has all finite weights 
$$
w_{\cal A}[\Omega,\Omega']\approx \delta[\tilde{\Omega},\tilde{\Omega}' ]
\exp\left(-{\bf i} 2 S \alpha \sum_{i}  \vec{n}\cdot\vec{n}_{\varphi_i}   \right)
$$
where $\tilde{\Omega}$ and $\tilde{\Omega}'$ means as representatives of the equivalence classes defined by ${\cal R}_{\Omega}^\dagger{\cal R}_{\Omega'}={\cal R}_{\alpha,\vec{n}_{\varphi}}$.

In this way, the integral in (\ref{eq:SRint1}) is reduced to an integral over every pair of rotations which differs in a 
rotation around the polarization axes. It can be decomposed as:
$$
\rho^{\rm SR'}_{\cal A}\rightarrow \rho^{\rm SR}_{\cal A} + \int {\bf R}^{\cal A}_{\Omega}
\Delta \rho_{\cal A} ({\bf R}^{\cal A}_{\Omega})^{\dagger} d\mu_{\Omega}\,.
$$
where 
$$
\Delta\rho_{\cal A}\propto \left (\int 
e^{-{\bf i} 2 S \alpha \sum_{i}  \vec{n}\cdot\vec{n}_{\varphi_i}}
{\bf R}_{\alpha,\vec{n}_{\varphi}}^{\cal A}\frac{d\alpha}{2\pi} \right)
\rho^{\rm Neel}_{\cal A}  + h.c.
$$
If $\sum_{i}  \vec{n}\cdot\vec{n}_{\varphi_i}=0$, $\Delta\rho_{\cal A}\propto \rho^{\rm SR}_{\cal A}$ and the result does not change. On the other hand, if $\sum_{i}  \vec{n}\cdot\vec{n}_{\varphi_i}=\pm 1$, 
the parenthesis results proportional to the projector over the local $({\bf S}^z)^{total({\cal A})}=\pm 2 S$. In a global N\'eel state, this projector is orthogonal to any subspace and then the contribution to the local state vanishes.


\vspace{1cm}

\bibliographystyle{unsrt-mine} 
\bibliography{biblio}
\end{document}